\documentclass[a4paper,11pt]{article}
\usepackage{graphicx}
\usepackage{epstopdf}
\usepackage{amsfonts,amsmath,amssymb}
\usepackage{hyperref}
\usepackage{float}

\usepackage{epsfig,multicol,bbm}

\newcommand\fverb{\setbox\pippobox=\hbox\bgroup\verb}
\newcommand\fverbit{\egroup\item[\fbox{\unhbox\pippobox}]}

\newbox\pippobox

\begin{document}
\title{\bf  From Taub-NUT to Kaluza-Klein magnetic monopole}
\author{Nematollah Riazi\thanks{Electronic address: n\_riazi@sbu.ac.ir},  S. Sedigheh Hashemi
\\
\small Department of Physics, Shahid Beheshti University, G.C., Evin, Tehran 19839,  Iran}
\maketitle
\begin{abstract}
We present a Kaluza-Klien vacuum solution which closely resembles
the Taub-NUT magnetic monopole and we investigate its physical
properties as viewed from four space-time dimensions. We show that
the Taub-NUT Kaluza-Klein vacuum solution in five dimensions is a
static magnetic monopole. We find that the four dimensional matter
properties do not obey the equation of state of radiation and
there is no event horizon. A comparison with the available
magnetic monopole solutions and the issue of  vanishing and negative mass are
discussed.
\end{abstract}


\section{Introduction}	
One of the oldest ideas that unify gravity and electromagnetism is
the theory of Kaluza and Klein which extends space-time to five
dimensions. The physical motivation for this unification
is that the vacuum solutions of the $(4+1)$ Kaluza-Klein field
equations reduce to the $(3+1)$ Einstein field equations with
effective matter and the curvature in $(4+1)$ space induces matter
in $(3+1)$ dimensional space-time\cite{1}.  With this idea, the
four dimensional energy-momentum tensor is derived from the
geometry of an exact five dimensional vacuum solution, and the
properties of matter such as density and pressure  as well as
electromagnetic properties are determined by such a solution. In
other words, the field equations of both electromagnetism and
gravity can be obtained from the pure five-dimensional geometry.

Kaluza\rq{}s idea was that the universe has four spatial
dimensions, and the extra dimension is compactified  to form a
circle so small as to be unobservable. Klein's contribution was to
make a reasonable physical basis for the  compactification  of the
fifth dimension. 
This school of thinking later led to
the eleven-dimensional supergravity theories in  1980s and to the
\lq\lq{}theory of everything\rq\rq{} or ten-dimensional
superstrings.

In this paper, we present a  vacuum solution of Kaluza-Klein
theory in five-dimensional spacetime which is closely related to
the Taub-NUT metric. The Taub-NUT  solution has many interesting
features; it carries a new type of charge (NUT charge), which has
topological origins and can be regarded as \lq\lq{}gravitational
magnetic charge\rq\rq{}, so the solution  is known in some other
contexts as the Kaluza-Klein magnetic monopole \cite{2}.

The plan of this paper is as follows. In section $2$,  we briefly
discuss  the formalism of five dimensional  Kaluza-Klein theory
and the effective four-dimensional Einstein-Maxwell equations. We
will then present a Taub-NUT-like Kaluza-Klien solution and
investigate its physical properties in four dimensions  in section 3. In the
last section we will draw our main conclusions.

 \section{Kaluza-Klein Theory }\label{sec2}
Kaluza $(1921)$ and Klein $(1926)$, used one extra dimension to
unify gravity and electromagnetism in a theory which was basically
five-dimensional general relativity. 
The theoretical
elegance of this idea is revealed by studying the vacuum solutions
of Kaluza-Klein equations and the matter induced in the
four-dimensional spacetime \cite{3}.Thus, we are chiefly
interested in the vacuum five dimensional Einstein equations.
 For any vacuum solution, the  energy-momentum tensor vanishes and thus $\hat G_{AB}=0$
or, equivalently $ \hat R_{AB}=0 $, where $\hat G_{AB} \equiv \hat
R_{AB}- \frac{1}{2}\hat R \hat g_{AB}$ is the Einstein tensor,
$\hat R_{AB}$ and $\hat R = \hat g_{AB} \hat R^{AB}$ are  the
five-dimensional Ricci tensor and scalar, respectively.  $\hat
g_{AB}$ is the metric tensor in  five dimensions. Here, the
indices $A,B,...$ run over $0...4$.

Generally, one can identify the $\mu \nu$ part  of $\hat g^{AB}$
with $g^{\mu \nu}$, which is the contravariant four dimensional
metric tensor, $A_{\mu}$ with the electromagnetic potential  and
$\phi$ as a scalar field.  The correspondence between the above
components is
\begin{equation}\label{eq1}
\hat g^{AB}= \left(
‎\begin{array}{ccccccc}‎
‎g^{\mu \nu}‎   & -\kappa A^{\mu} &  \\‎
 ‎‎  &‎\\‎
-\kappa A^{\nu}    ~~& \kappa^2 A^{\sigma}A_{\sigma}+\phi^2 &  \\‎
‎\end{array}
‎\right)
\end{equation}
where $\kappa$ is a coupling  constant for  the electromagnetic potential $A^{\mu}$
and the indices $\mu,\nu$ run over $0...3$.

The five dimensional field equations reduce to the four dimensional field equations\cite{4}
\begin{align}\label{eq2}
G_{\mu \nu}=\frac{\kappa^2 \phi^2}{2}T_{\mu \nu}^{EM}-\frac{1}{\phi}\left[\nabla_{\mu}(\partial _{\nu} \phi)
-g_{\mu \nu}\Box \phi\right],
\end{align}
and
\begin{align}\label{eq3}
\nabla^{\mu}F_{\mu \nu}=-3\frac{\partial ^{\mu}\phi}{\phi}F_{\mu \nu},  \nonumber\\
\Box \phi= \frac{\kappa^2 \phi^3}{4} F_{\mu \nu} F^{\mu \nu},
\end{align}
where $G_{\mu \nu}$ is the Einstein tensor,
$T_{\mu \nu}^{EM}\equiv \frac{1}{4} g_{\mu \nu}F_{\rho \sigma}F^{\rho \sigma}-F_{\mu}^{~\sigma}F_{\nu \sigma}$
is the electromagnetic energy-momentum tensor
 and  the field strength $F_{\mu\nu}=\partial_{\mu} A_{\nu}-\partial_{\nu}A_{\mu}$. Knowing the five dimensional
 metric, therefore, leads to a complete knowledge of the four dimensional geometry, as well as the electromagnetic
 and scalar fields.
\section{ Taub-NUT Solution And Kaluza-Klein Magnetic Monopole}
The Taub-NUT solution was first discovered by Taub (1951), and
subsequently by Newman, Tamburino and Unti (1963) as a
generalization of the Schwarzschild spacetime \cite{5}.

The Kaluza-Klein monopole of Gross and Perry is  described by the
following metric which is a generalization of the self-dual
Euclidean Taub-NUT solution \cite{6}
\begin{align}\label{eq4}
 {\rm d}s^2=- {\rm d}t^2+V( {\rm d}x^5+4m(1-\cos\theta) {\rm d}\phi)^2+\frac{1}{V}( {\rm d}r^2+r^2 {\rm d} \theta^2 +r^2 \sin ^2\theta  {\rm d}\phi^2),
\end{align}
where
\begin{align}
\frac{1}{V}=&1+\frac{4m}{r}.
\end{align}
This solution has a coordinate singularity at $r=0$ which is
called NUT singularity. This can be absent if the coordinate $x^5$
is periodic with period $16\pi m=2\pi R$, where $R$ is the radius
of the fifth dimension.
Gross and Perry showed that the Kaluza-Klein theory can contain
magnetic monopole solitons which would support the unified gauge
theories and allow us for searching the physics of unification.

Here, we  introduce a metric which is a vacuum five dimensional
solution, having some properties in common with the monopole of
Gross and Perry, while some other properties being different. The
proposed static metric is given by
\begin{align}\label{6}
 {\rm d}s^2_{(5)}= - {\rm d}t^2+w(r)\left( {\rm d}r^2+r^2 {\rm d}\theta^2 +r^2\sin^2 \theta  {\rm d}\phi^2\right)+
 \frac{k}{w(r)}\left( {\rm d}\psi + Q\cos\theta  {\rm d}\phi\right)^2.
\end{align}
Here, the extra coordinate is represented by $\psi$.   $k$,  $Q$ are constants and $w(r)$ is an arbitrary function of the radial coordinate $r$. The coordinates take on the usual values with range
$r\geq0$, $0\leq\theta\leq\pi$, $0\leq  \phi\leq 2\pi$ and $0\leq
\psi \leq 2\pi$. 

The Ricci scalar associated with the five dimensional metric (\ref{6}) is given by
\begin{equation}\label{7}
R=\frac{1}{2}\frac{1}{w(r)^3r^4}\left[2r^4w(r)w''(r)+4w\rq{}(r)w(r)r^3 -w\rq{}(r)^2 r^4+kQ^2\right],
\end{equation}
 since we are interested in vacuum solution where the Ricci tensor is zero $(R_{AB}=0)$ it is necessary but not sufficient to have a zero Ricci scalar $R=0$ which can be solved for the function $w(r)$ to give
\begin{equation}
w(r)=k_1+\frac{k_2}{r}+\frac{k_3}{r^2},
\end{equation}
 where $k_1$, $k_2$ and $k_3$ are constants. By substituting  the function $w(r)$ in the Ricci scalar in Eq.~(\ref{7}) and for having a zero Ricci scalar one can obtain the following constraint between the constants
\begin{equation}
kQ^2+4k_1k_3-k_2^2=0,
\end{equation}
or
\begin{equation}
k_3=\frac{1}{4k_1}\left(k_2^2-kQ^2\right),
\end{equation}
 thus the Ricci scalar in terms of the above constants reduces to
\begin{equation}
R=\frac{1}{2}\frac{r^2(4k_1-1)(kQ^2-k_2^2)}{(kQ^2-k_2^2-k_2r-k_1r^2)^3}.
\end{equation}
In order to have a zero Ricci scalar the two following  possibilities can be obtained for the constants
 \begin{equation}\label{12}
R=0 \Longrightarrow \begin{cases}
k_1=\frac{1}{4} \Longrightarrow R_{AB}\neq 0 ,\\
k_2=\pm \sqrt {k}Q \Longrightarrow R_{AB}= 0, \quad R_{ABCD}\neq0,
\end{cases}
\end{equation}
 as it is seen from Eq.~(\ref{13}) with $k_1=\frac{1}{4}$ the Ricci tensor is not zero and does not give a vacuum solution, thus this case is discarded. However, the second choice $k_2=\pm \sqrt {k}Q$ gives a  Ricci flat solution with a non-zero Riemann tensor in five dimensions which is our interest. Consequently, the general form of the function $w(r)$
would be given by
\begin{equation}
\label{13}
w(r)=k_1\pm \frac{\sqrt {k}Q }{r}.
\end{equation}
 In what follows, we take the constants $k_1=1$, $Q=1$ and $k=4m^2$. Therefore, the metric in Eq.~(\ref{6})  reduces to (we choose the minus sign for $w(r)$)
\begin{align}\label{14}
{\rm d}s^2_{(5)}&= - {\rm d}t^2+(1-\frac{2m}{r})\left( {\rm d}r^2+r^2 {\rm d}\theta^2 +r^2\sin^2 \theta  {\rm d}\phi^2\right) \nonumber\\ &
+\left(\frac{4m^2}{1-\frac{2m}{r}}\right)\left( {\rm d}\psi + Q\cos\theta  {\rm d}\phi\right)^2.
\end{align}

The gauge field, $A_{\mu}$ and the scalar field
$\phi$ deduced from the metric in Eq.~(\ref{14}) with the help of Eq.~(\ref{eq1})  are
\begin{equation}
A_{\phi}=\frac{\cos\theta}{\kappa},
 \end{equation}
 and
 \begin{equation}
 \phi^2=\frac{4m^2}{1-\frac{2m}{r}},
 \end{equation}
respectively.
 The only non-vanishing component for the electromagnetic tensor which is related to the gauge field $A_\mu$ by the general statement  $F_{\mu \nu}=\partial _\mu A_\nu-\partial _\nu A_\mu$ is
\begin{equation}
F_{\theta \phi}=-F_{\phi \theta}= -\frac{\sin \theta}{\kappa},
\end{equation}
which corresponds to a radial magnetic field  $B_{r}=\dfrac{1}{\kappa r^2}$ with a magnetic monopole 
charge $Q_{M}=\dfrac{1}{\kappa}$. As the radial coordinate  $r$ goes to  infinity $r\rightarrow \infty$, the scalar field
equals $\phi_{0}^2=4m^2$   so that the
second part of Eq.~(\ref{eq2}) becomes zero, therefore
\begin{equation}
G_{\mu \nu}=\frac{\kappa^2\phi_{0}^2}{2}T^{EM}_{\mu \nu}=8\pi GT_{\mu \nu}, \qquad as\qquad r\rightarrow \infty,
\end{equation}
where we have put the speed of light $c$ equal to $1$. By
comparing this equation with the ordinary Einstein equation, we
obtain $\dfrac{\kappa^2 \phi_{0}^2}{2}=8\pi G$, thus the constant
$\kappa$ equals $\kappa=\dfrac{2}{m}\sqrt{\pi G}$.  Straightforwardly, the total magnetic monopole  charge is given by
\begin{equation}\label{eq11}
Q_{M}=\frac{m}{2}\frac{1}{\sqrt{ \pi G}}.
\end{equation}
The total magnetic flux through any spherical surface centered at
the origin is calculated as\cite{7}
\begin{equation}\label{22}
\Phi_{B}=\int F=\frac{1}{2}\oint  F_{\mu\nu}d\Sigma^{\mu
\nu}=\frac{\pi}{\kappa},
\end{equation}
 where $F_{\mu \nu}$ is the electromagnetic field tensor as mentioned earlier and ${\rm d}\Sigma^{\mu \nu}$ is interpreted as an element of two-dimensional surface area.
It is investigated from Eq.~(\ref{22}) that  the flux of the magnetic monopole is constant
(i.e. we have a singular magnetic charge).

The four dimensional metric deduced  from Eq.~(\ref{14}) with the use of Eq.~(\ref{eq1}) leads to the following asymptotically flat spacetime:
\begin{align}\label{26}
 {\rm d}s^2_{\left(4\right)}=- {\rm d}t^2+\left(1-\frac{2m}{r}\right) {\rm d}r^2+r^2\left(1-\frac{2m}{r}\right)\left( {\rm d}\theta^2 +\sin ^2\theta 
 {\rm d}\phi^2\right).
\end{align}
 For this metric,
the Ricci scalar  and the Kretschmann  invariant $K$ are given by
\begin{equation}
R=\frac{6m^2}{r^4\left(1-\frac{2m}{r}\right)^3},
\end{equation}
and
 \begin{equation}
 K=R_{\mu \nu \rho \sigma}R^{\mu \nu \rho \sigma}=\frac{m^2}{\left(r-2m\right)^5}\left[\frac{2}{r-2m}+\frac{2r-3m}{r^2}\right],
 \end{equation}
 it is seen that the four dimensional metric Eq.~(\ref{26}) has two curvature singularities at $r=0$ and $r=2m$.
  Let us calculate the surface area of a $S^2$ hypersurface at constant $t$
and $r$ to clarify the nature of the $r=2m$ singularity
\begin{equation}
A\left(r\right)=\int \sqrt{g^{\left(2\right)}} {\rm d}x^2= \int r^2\left(1-\frac{2m}{r} \right)\sin \theta  {\rm d} \theta  {\rm d} \phi=
4 \pi  r^2\left(1-\frac{2m}{r} \right).
\end{equation}
It can be seen that the surface area $A\left(r\right)$ becomes
zero at $r=2m$. This means that $r=2m$
  hypersurface is not an event horizon but is just the origin.
For the case $r>2m$ the signature of the metric is proper $(-,+,+,+)$
but for the range $r<2m$ the signature of
 the metric will be improper
and non-Lorentzian $ (-,-,-,-)$, thus the patch
$r<2m$ is excluded from the spacetime. From now  on,  this spacetime is
only considered in the range  $r\geq2m$.
 since the range $0<r<2m$ is omitted from the spacetime, there is only one curvature singularity at
 $r=2m$.

The metric (\ref{26}) can be transformed into the following form,
using the radial coordinate transformation ${\tilde r}^2=r^2\left(
1-\frac{2m}{r}\right)$;
\begin{equation}
 {\rm d}s^2_{(4)}=- {\rm d}t^2+\frac{{\tilde r}^4}{(m^2+{\tilde r}^2)(m+
\sqrt{m^2+{\tilde r}^2})^2} {\rm d}{\tilde r}^2 +{\tilde r}^2 {\rm d}\Omega^2.
\end{equation}
While the  metric (\ref{26}) is singular at  $r=0$ and $r=2m$,
the transformed metric is singular only at ${\tilde r}=0$ which is a
curvature singularity.

The properties of the induced matter associated with the above
metric can be gained by the Eq.~(\ref{eq3}).
The components of the energy-momentum tensor can be easily
calculated for the metric (\ref{26}). We find that the effective
source is like a fluid with an anisotropic pressure. At
sufficiently large $r$, the energy-momentum components will tend
to zero. The trace of the energy-momentum tensor does not vanish
generally, which shows that the effective matter field around the
singularity can not be considered as an ultra-relativistic quantum
field in contrast to the Kaluza-Klein solitons described in Ref. cite{8}. 

Although it is apparent from the $00$  component of the $4D$ metric  (\ref{26}) that the gravitational mass corresponding to this solutions vanishes, it could also be confirmed by calculating the following integrals:
\begin{equation}
M_{g}\left(r\right) \equiv \int \left(T^{0} _{~0}-T^{1} _{~1}-T^{2} _{~2}-T^{3} _{~3} \right)\sqrt{g^{(3)}} {\rm d}^{3}x,
\end{equation}
and
\begin{equation}\label{eq29}
M_{g}(r)=-\frac{1}{8\pi}\oint _{S}\nabla ^{\mu}K^{\nu} {\rm d}S_{\mu \nu},
\end{equation}
where ${g^{(3)}}$ is the determinant of the 3-metric.

\section{Conclusion}
Inspired by  the Taub-NUT solution, we   introduced a Kaluza-Klein
vacuum solution in $5D$, which described a magnetic monopole in
$(3+1)$D spacetime. The gravitational mass of the solution was
shown to vanish. The fluid supporting the four dimensional
space-time was shown to differ from that of an ultra-relativistic
fluid, in contrast with the work by Wesson and Leon \cite{8}. The
pressure is anisotropic in both works.

\section*{Acknowledgments}
The authors would like to thank the anonymous referee for
helpful comments. N.R. Acknowledges the support of Shahid Beheshti University.


\begin{thebibliography}{0}    


\bibitem{1}
J. Ponce de Leon and P. S. Wesson, {\it Journal of Math. Phys}. {\bf 34},
4080-4092 (1993).
\bibitem{2}
T. Ortin, {\it Gravity and String}, (Cambridge University Press, 2004).
 \bibitem{3}
P. S. Wesson and J. Ponce de Leon.  {\it Journal of Math. Phys}. {\bf 33},
3883 (1992).
\bibitem{4}
G. Lessner, {\it Phys. Rev. D}. {\bf 25}, 3202 (1982).
 \bibitem{5} 
 D. Baleanu and S. Codoban, {\it General Relativity and Gravitation}. {\bf 31}, 497  (1999).
 \bibitem{6} 
 D. J. Gross  and M. J. Perry, {\it  Nucl. Phys. B}. {\bf 29},  226 (1983).
 \bibitem{7}
 R. D. Sorkin, {\it  Phys. Rev. Lett}. {\bf  51}, 87 (1983).
 \bibitem{8}
Paul S. Wesson and J. Ponce de Leon.  {\it Classical and Quantum
Gravity}. {\bf  11}, 1341 (1994).
e de Leon. {\it Astronomy and Astrophysics}. {\bf 1}, 294 (1995).

\end{thebibliography}
\end{document}